\documentclass[12pt]{article}
\usepackage{amssymb,amsfonts}
\newlength{\extraspace}
\newlength{\extraspaces}
\newcommand{\beq}{\begin{eqnarray}
\addtolength{\abovedisplayskip}{\extraspaces}
\addtolength{\belowdisplayskip}{\extraspaces}
\addtolength{\abovedisplayshortskip}{\extraspace}
\addtolength{\belowdisplayshortskip}{\extraspace}}
\newcommand{\eeq}{\end{eqnarray}}

\newcommand{\newsection}[1]{
\vspace{15mm}
\pagebreak[3]
\addtocounter{section}{1}
\setcounter{equation}{0}
\setcounter{subsection}{0}
\setcounter{footnote}{0}
\begin{flushleft}
{\large\bf \thesection. #1}
\end{flushleft}
\nopagebreak
\medskip
\nopagebreak}

\newcommand{\newsubsection}[1]{
\vspace{1cm}
\pagebreak[3]
\addtocounter{subsection}{1}
\noindent{ \bf \thesubsection. #1}
\nopagebreak
\vspace{2mm}
\nopagebreak}

\begin{document}

\addtolength{\baselineskip}{.8mm}

{\thispagestyle{empty}
\noindent \hspace{1cm}  \hfill GEF--TH/2003--2 \hspace{1cm}\\

\begin{center}
\vspace*{1.0cm}
{\large\bf Field theoretical approach to the study of theta dependence} \\
{\large\bf in Yang-Mills theories on the lattice} \\
\vspace*{1.0cm}
{\large Massimo D'Elia}\\
\vspace*{0.5cm}{\normalsize
{Dipartimento di Fisica dell'Universit\`a di Genova and I.N.F.N., \\
Via Dodecaneso 33, I-16146 Genova, Italy.}} \\
\vspace*{2cm}{\large \bf Abstract}
\end{center}

\noindent
We discuss the extension of the field theoretical approach, already 
used in the lattice determination of the topological susceptibility,
to the computation of further terms in the expansion of the ground 
state energy $F(\theta)$ around $\theta = 0$ in $SU(N)$ Yang-Mills 
theories. In particular we determine the fourth order term in the 
expansion for $SU(3)$ pure gauge theory and compare our results with 
previous cooling determinations. In the last part of the paper we make 
some considerations about the nature of the ultraviolet fluctuations 
responsible for the renormalization of the lattice topological charge 
correlation functions; in particular we propose and test an ansatz 
which leads to improved estimates of the fourth and higher order terms 
in the expansion of $F(\theta)$.

\vspace{0.5cm}
\noindent
(PACS Numbers: 11.15.Ha, 11.10.Gh, 12.38.Gc)
}
\vfill\eject

\newsection{Introduction}              

The dependence of $SU(N)$ Yang--Mills theories on the $\theta$ angle
is the subject of ongoing  theoretical debate. The dependence
can be expressed in terms of the free energy density $F(\theta)$
which, in the Euclidean theory, is defined as follows:
\beq
\exp[ - V F(\theta) ] \equiv 
Z(\theta) = \int [dA] \; e^{ - \int d^4 x {\cal L}(x)} \;  
 e^{ i \theta\, Q} \; ,
\label{fthetadef}
\eeq
where $V$ is the four dimensional volume, 
${\cal L}(x) = {1\over 4} F_{\mu\nu}^a(x)F_{\mu\nu}^a(x)$ 
is the usual Yang-Mills langrangian and 
$Q = \int \hbox{d}^4x \; q(x)$ is the topological charge, with the topological
charge density $q(x)$ defined as
\beq
q(x) = {g^2\over 64\pi^2} \epsilon_{\mu\nu\rho\sigma}
F_{\mu\nu}^a(x)F_{\rho\sigma}^a(x) = \partial_\mu K_\mu(x)\; ,
\label{defqx}
\eeq
where $K_\mu(x)$ is the Chern current 
\beq
 K_\mu = {g^2 \over {16 \pi^2}} \epsilon_{\mu\nu\rho\sigma}
 A^a_\nu \left(\partial_\rho A^a_\sigma - {1 \over 3}
 g f^{abc} A^b_\rho A^c_\sigma \right) \; .
\label{eq:k}
\eeq

The coefficients of the Taylor expansion of $F(\theta)$ around 
$\theta = 0$, 
\beq
F(\theta) = \sum_{k = 0}^\infty \frac{1}{k!} F^{(k)}(0) \theta^k \; ,
\label{fthetaexp}
\eeq
are related to the connected expectation values of the topological
charge distribution,
\beq
 F^{(k)}(0) \equiv \frac{d^k}{d \theta^k} F(\theta) |_{\theta=0} = 
 - i^k \frac{\langle Q^k \rangle_c}{V} \; ,
\label{expcoeff}
\eeq
where $\langle \cdot \rangle_c$ is a short notation meaning the connected 
expectation value taken at $\theta = 0$.

$F(\theta)$ is a non-trivial function, indeed the topological susceptibility,
$\chi = \langle Q^2 \rangle_c/V$, is expected
to be different from zero, to the leading
order in $1/N$, to solve the so--called $U(1)$ 
problem~\cite{wit-79,ven-79}.
In Refs.~\cite{wit-80,wit-98} it has been argued that, in the large $N$ limit,
$F(\theta)$ is a multibranched function of $\theta$, and in particular
\beq
F(\theta) = F(0) + \frac{\chi}{2} \, {\rm min}_k \, (\theta+2\pi k)^2 + O\left(
1/N\right) \; .
\label{wit-conj}
\eeq 
Therefore, for sufficiently small values of $\theta$
($\theta < \pi$),
$F(\theta)$ is expected to have an almost quadratic dependence
on $\theta$, with $O(\theta^4)$ corrections suppressed by powers of $1/N$.

Numerical  Monte Carlo simulations on the lattice are
a natural tool to obtain information from first principles about the 
dependence of $F(\theta)$ around $\theta = 0$.
While direct numerical simulations of 
the theory at $\theta \neq 0$ are  not feasible because of the 
complex nature of the action, the Taylor expansion of 
$F(\theta)$ around $\theta = 0$ can be computed, in principle
up to any given order, by measuring 
the connected expectation values of the topological charge 
over the ensemble of configurations at $\theta = 0$, as explicited in 
Eqs.~(\ref{fthetaexp}) and~(\ref{expcoeff}).
The topological susceptibility
has been already extensively studied on the lattice 
(see Refs.~\cite{tep-00,gar-01} for recent reviews)
and further terms in the expansion have been recently measured~\cite{ehl-02}.

The lattice study of quantities related to topology requires
care. The problem is usually related to the fact that
the topology of gauge configurations on a discretized space-time 
is, strictly speaking, always trivial, and that the usual lattice definition,
given in terms of gauge fields as a na\"{\i}ve discretization of the continuum
topological charge, does not have the continuum integer valued spectrum:
as a good alternative the fermionic definition, which is directly 
related to the index theorem, or the definition based on the so-called
cooling method, are used.

In fact, the topological charge operator and its correlation functions
can be defined on the lattice with the same rigour as for any other
operator of the theory: as for any other physical quantity, one has 
to pay attention when removing the ultraviolet (UV) regulator, 
i.e. when going to the continuum limit, since the appropriate 
renormalizations have to be performed.
In the cooling method the UV lattice fluctuations, which are responsible
for the renormalizations, are removed by a process of local 
minimization of the action. 
However it is also possible to compute the renormalizations and perform
the appropriate subtractions. This program, usually known as
the ``field theoretical method'', has been already widely discussed and 
developed, together with a method for the numerical non-perturbative  
determination of the renormalization constants, usually known
as the ``heating method'', in the context of the lattice determination
of the topological 
susceptibility~\cite{cdp-88,teper-89,cdpv-90,dv-92,acdgv-93,fp-94,add-97,add-97-2,addk-98}.

The aim of the present paper is that of discussing the extension 
of the field theoretical method (and of the heating method
used to compute the renormalizations) to the case of higher order 
correlation functions of the topological charge, in order to study
the $\theta$ dependence of the theory.
In Section 2, after a review of the field theoretical method,
as used for the computation of the topological susceptibility, 
we will discuss its application to the case of higher order correlations
and develop a suitable extension of the heating method.
In Section 3 the case of 
$SU(3)$ pure gauge theory will be used as a testground for 
the method developed in Section 2, and we will determine the fourth
order contribution to $F(\theta)$ and compare our results
with those obtained by the cooling technique~\cite{ehl-02}.
In Section 4 we will state and test an ansatz about the nature
of the UV lattice fluctuations responsible for the renormalizations, which
will allow us to simplify the computation of the 
connected correlation functions and to obtain more precise 
determinations of the fourth order contribution to $F(\theta)$.
Finally, in Section 5, we will give our conclusions.

\newsection{Topological charge correlation functions on the lattice}

In this Section we will discuss how the various moments of the lattice
topological charge distribution renormalize with
respect to the continuum ones, and how the corresponding renormalizations 
can be computed numerically.
In order to make the discussion clearer, we will first review 
the case of the second moment, i.e. the topological susceptibility.

\newsubsection{Renormalization of the topological susceptibility}

On the lattice it is possible to define a discretized gauge invariant 
topological charge density operator $q_L(x)$, and a related
topological charge $Q_L = \sum_x q_L(x)$ (with the sum extended over
all lattice points), with the only requirement that, 
in the formal (na\"{\i}ve) continuum limit, 
\beq
 q_L(x) {\buildrel {a \rightarrow 0} \over \sim} a^4 q(x) + O(a^6) \; ,
\label{eq:naive}
\eeq
where $a$ is the lattice spacing.
A possible definition is
\beq
q_L(x) = {{-1} \over {2^9 \pi^2}} 
\sum_{\mu\nu\rho\sigma = \pm 1}^{\pm 4} 
{\tilde{\epsilon}}_{\mu\nu\rho\sigma} \hbox{Tr} \left( 
\Pi_{\mu\nu}(x) \Pi_{\rho\sigma}(x) \right) \; ,
\label{eq:qlattice}
\eeq
where $\Pi_{\mu\nu}(x)$ is the usual plaquette operator in the 
$\mu\nu$ plane, ${\tilde{\epsilon}}_{\mu\nu\rho\sigma}$ is the
standard Levi--Civita tensor for positive directions  and is otherwise
defined by the rule ${\tilde{\epsilon}}_{\mu\nu\rho\sigma} =
- {\tilde{\epsilon}}_{(-\mu)\nu\rho\sigma}$.

A proper renormalization must be performed when going towards 
the continuum limit, like for any other regularized operator. 
In spite of the  formal limit
in Eq.~(\ref{eq:naive}), the discretized topological charge
density renormalizes  multiplicatively~\cite{cdp-88}:
\beq
 q_L(x) = Z(\beta) a^4(\beta) q(x) + O(a^6) \; , 
\eeq
with a multiplicative renormalization constant $Z(\beta)$ which is a finite 
function of the bare coupling $\beta = 2 N / g_0^2$, approaching
1 as $\beta \to \infty$. 

When defining the topological susceptibility,
further renormalizations  can appear.
Indeed, already the continuum definition,
\beq
\chi \equiv \frac{\langle Q^2 \rangle}{V} = 
\int \hbox{d}^4x \; \langle q(x) q(0)  \rangle \; ,
\label{eq:chi}
\eeq
involves the product
of two operators $q(x)$ at the same point: this contact term is divergent
and not well defined, so that an appropriate prescription must be assigned.
It can be shown~\cite{crew-79,meg-98} that the correct prescription, corresponding
to the quantity which appears in the Taylor expansion of $F(\theta)$,
is the one in which the derivative appearing in the definition
of $q(x)$, Eq. (\ref{defqx}), is taken out of the vacuum expectation value:
\beq
 \chi = \frac{1}{V} \int \hbox{d}^4 x \hbox{d}^4 y \; \partial^x_\mu  \partial^y_\nu
\langle K_\mu(x) K_\nu(y) \rangle \; .
\label{eq:prescription}
\eeq
The lattice definition of the topological susceptibility
\beq
\chi_L = \sum_x \langle  q_L(x) q_L(0) \rangle 
\label{lat_chi}
\eeq
is in general not guaranteed to meet the correct continuum prescription
for the contact term, and this leads to the appearance of additive
renormalizations:
\beq
\chi_L = Z(\beta)^2 a^4(\beta) \chi + M(\beta) \; ,
\label{renchi}
\eeq
where $M(\beta)$ describes generically the mixing
with all local scalar operators appearing
in the operator product expansion (OPE) of 
$q_L(x) q_L(0)$ as $x \sim 0$ in Eq. (\ref{lat_chi}), including in 
particular the action density and the identity operator.

The idea behind the numerical technique, known as the heating method, 
used to compute the two renormalizations $Z(\beta)$ and $M(\beta)$, 
is that the UV fluctuations in $q_L(x)$, which are responsible for
renormalizations, are decoupled from the background topological
signal so that, starting from a semiclassical configuration of fixed
and well known topological content, it is possible, by applying
a few thermalization steps (i.e. Monte Carlo updating steps at the 
corresponding value of $\beta$),
to thermalize the UV fluctuations without altering the
background topological content. This is certainly true for 
high enough $\beta$, i.e. approaching the continuum limit, 
and in practice it turns out to be true in a range of the $\beta$ values 
usually chosen in Monte Carlo simulations of Yang-Mills theory,
being also favoured by the fact that topological modes have very 
large autocorrelation times, as compared to any other non-topological mode.

It is thus possible to create samples 
of configurations which have a fixed topological content, $Q$, and 
the UV fluctuations thermalized:
various measurements of topological quantities on these samples can give
information about the renormalizations. For example
the expectation value of $Q_L$ gives
\beq
\langle Q_L \rangle = Z(\beta) \;  Q
\label{heat-z}
\eeq
from which the value of $Z(\beta)$ can be inferred, while the expectation
value of $\langle Q_L^2 \rangle$ gives 
\beq
\langle Q_L^2 \rangle = Z(\beta) \; Q^2 + V \; M(\beta) \; ,
\label{heat-0}
\eeq
where by $V$ we intend, from now on, the four dimensional volume 
measured in adimensional lattice units.

To check that UV fluctuations have been thermalized, one looks
for plateaux in quantities like $\langle Q_L \rangle$ or 
$\langle Q_L^2 \rangle$ as a function of the heating steps performed:
only configurations obtained after the plateau has been reached are 
included in the sample. Special care has to be paid to verify
that during the heating procedure the background
topological charge is left unchanged: this is usually done
by performing a few cooling steps on a copy of the heated configuration
and configurations where the background topological content 
has changed are discarded from the sample~\cite{fp-94}. Further details
about the procedure and about the estimate of the systematic
errors involved can be found in Ref.~\cite{add-97}.

A sample with $Q \simeq 1 $ can be used to measure $Z$
and a sample with $Q = 0$ (usually thermalized around
the zero field configuration) can be used to determine $M$.
Cross-checks can then be performed, using samples obtained starting
from various semiclassical configurations with the same 
or different values of $Q$, to test the validity of the 
method.

Once the renormalizations have been computed and the 
expectation value $\chi_L$ over the equilibrium ensemble has been
measured, the physical topological susceptibility $\chi$ can be extracted,
using Eq.~(\ref{renchi}), as
\beq
\chi = \frac{\chi_L - M(\beta)}{a^4(\beta) Z(\beta)^2}
\label{subchi}
\eeq
\\

If large renormalization effects are present, i.e. if $Z~\ll~1$
and if $M$ brings a good fraction of the whole signal in $\chi_L$, 
the final determination of $\chi$, obtained via Eq.~(\ref{subchi}),
can be affected by large error bars. However one can exploit
the fact that $Z$ and $M$ both depend on the lattice discretization
$q_L(x)$ and that
infinitely many operators $q_L(x)$ can be defined all having the same
na\"{\i}ve continuum limit, to choose improved operators
for which the renormalization effects are reduced, thus leading to
improved estimates of $\chi$. This is the idea behind the definition
of smeared operators~\cite{cdpv-96}
\beq
q_L^{(i)}(x) = {{-1} \over {2^9 \pi^2}} 
\sum_{\mu\nu\rho\sigma = \pm 1}^{\pm 4} 
{\tilde{\epsilon}}_{\mu\nu\rho\sigma} \hbox{Tr} \left( 
\Pi_{\mu\nu}^{(i)}(x) \Pi_{\rho\sigma}^{(i)}(x) \right) \; ,
\label{qlsmear}
\eeq
where $\Pi_{\mu\nu}^{(i)}(x)$ is the plaquette operator constructed 
with $i$--times smeared links $U_\mu^{(i)}(x)$, which are defined
as 
\beq
U^{(0)}_{\mu}(x) &=& U_{\mu}(x) \; , \nonumber \\
{\overline U}^{(i)}_{\mu}(x) &=& (1-c) U^{(i-1)}_{\mu}(x) +
{c \over 6} 
\sum_{{\scriptstyle \alpha = \pm 1} \atop { \scriptstyle 
|\alpha| \not= \mu}}^{\pm 4}
U^{(i-1)}_{\alpha}(x) U^{(i-1)}_{\mu}(x+\hat{\alpha})
U^{(i-1)}_{\alpha}(x+\hat{\mu})^{\dag}, \nonumber \\
U^{(i)}_{\mu}(x) &=& {{{\overline U}^{(i)}_{\mu}(x)} \over
{ \left( {1 \over 3} \hbox{Tr} {\overline U}^{(i)}_{\mu}(x)^{\dag} 
{\overline U}^{(i)}_{\mu}(x) \right)^{1/2} } } \; ,
\eeq
where $c$ is a free parameter which can be tuned to optimize the improvement.
These operators have been successfully used, up to the second
smearing level, to determine $\chi$ at zero and finite
temperature both in $SU(2)$~\cite{add-97-2} (with $c = 0.85$) and 
$SU(3)$~\cite{add-97} (with $c = 0.9$) pure gauge theory. We will
make reference to them later in this paper, when computing the higher order
correlation functions of the topological charge.

\newsubsection{Renormalization of higher order correlation functions}

In order to compute the higher order connected moments of
the topological charge distribution, $\langle Q^n  \rangle_c$, needed for
the Taylor expansion of $F(\theta)$, it is necessary
to first compute the disconnected correlation functions
$\langle Q^n \rangle$. We will consider in particular
the case $n = 4$, for which  
$\langle Q^4  \rangle_c = \langle Q^4 \rangle - 3 \langle Q^2 \rangle^2 $.

Like in the case of the topological susceptibility, 
also the definition of $\langle Q^4 \rangle$ needs a special prescription
for the contact terms, in order to correspond to the quantity which enters
the Taylor expansion of $F(\theta)$.
Starting from the lattice definition, $Q_L$, one can define
the expectation value 
\beq
\langle Q_L^4 \rangle = 
\int \hbox{d}^4x_1 \dots \hbox{d}^4x_4
\langle q_L(x_1) q_L(x_2) q_L(x_3) q_L(x_4) 
\rangle
\label{mulcon}
\eeq
and it is clear that, apart from an  obvious multiplicative renormalization
constant $Z^4 (\beta)$, $\langle Q_L^4 \rangle$ will be linked to 
$\langle Q^4 \rangle$ by additive renormalizations related
to the contact terms arising when two or more charge densities in the 
expectation value integrated in Eq.~(\ref{mulcon}) come to the same point.
Our aim is to eventually compute 
the renormalization constants numerically, and in this sense
we are not interested in 
the exact form of the mixing terms, but just in understanding
if they can be related in a simple way  to
the correlation functions of different order, $\langle Q^m \rangle$,
so that they can be computed using 
the heating method, as it will be clarified below.

We will assume a simple and quite natural form for 
the renormalization rule: 
\beq
\langle Q_L^4 \rangle &=& Z(\beta)^4 \langle Q^4 \rangle + 
M_{4,2}(\beta) \langle Q^2 \rangle + M_{4,0}(\beta) \; ,
\label{ren-rule}
\eeq
where $M_{4,2}(\beta)$ and $M_{4,0}(\beta)$ are two constants which are 
independent of the topological sector. We will discuss 
in the following whether Eq. (\ref{ren-rule}) makes sense from a theoretical
point of view, while its validity will be fully checked numerically 
in Section 3.

The presence of the term $ Z(\beta)^4 \langle Q^4 \rangle $ in 
Eq. (\ref{ren-rule}) comes from the multiplicative
renormalization of the topological charge density. The question is then 
whether the additive renormalizations coming from contact terms
can be put in the form $M_{4,2}(\beta) \langle Q^2 \rangle + M_{4,0}(\beta)$.
Contact terms arise when two or more charge densities in the 
expectation value in Eq.~(\ref{mulcon}) come to the same point, but
the discussion of the two charge case will suffice:
the instance with three (or more) charge densities coming
to the same point can be considered as a special case of the 
two charge case, so that the related mixing terms are already considered
when discussing the two charge contact term.

Let us consider for instance $x_3 \sim x_4$ in Eq. (\ref{mulcon}). 
The operators of lower dimension appearing in the OPE of 
$q_L(x_3) q_L(x_4)$ are the identity operator and the 
action density $s(x) \propto F^a_{\mu\nu}(x) F^a_{\mu\nu}(x)$.
The insertion of the identity operator in the expectation value
in Eq. (\ref{mulcon}) leads to a contribution proportional to
\beq
&& \int \hbox{d}^4x_1 \hbox{d}^4x_2 \; 
\langle q_L(x_1) q_L(x_2) \; {\rm Id} \rangle =
 \langle Q_L^2 \rangle = Z^2(\beta) \langle Q^2 \rangle + V \; M(\beta) \; ,
\eeq
which is consistent with the assumption in Eq. (\ref{ren-rule}).
The insertion of the action density operator leads to a contribution 
proportional to
\beq
 \int \hbox{d}^4x_1 \hbox{d}^4x_2 \hbox{d}^4x 
\langle q_L(x_1) q_L(x_2) s(x) \rangle = \langle Q_L^2 S \rangle =
 \langle S \rangle \langle Q_L^2 \rangle - \frac{d}{d \beta} 
\langle Q_L^2 \rangle \; ,
\label{acmi}
\eeq
where $S$ is the total action of the theory and we have used the relation,
which is valid for any operator $O$, $(d/d\beta) \langle O \rangle = 
\langle S \rangle \langle O \rangle - \langle S O \rangle$,
where the derivative is taken as the four dimensional volume
in lattice units, $V$, is kept fixed.
The first term on the right hand side of Eq. (\ref{acmi}), 
$ \langle S \rangle \langle Q_L^2 \rangle$, is  proportional 
to $ Z^2(\beta) \langle Q^2 \rangle + V \; M(\beta) $ and therefore
consistent with Eq. (\ref{ren-rule}). Using
the fact that the topological susceptibility $\chi$ is a renormalization 
group invariant quantity, i.e. independent of $\beta$, 
and that $\chi = \langle Q^2 \rangle / (V a^4)$, the second
term can be written as
\beq
 \frac{d}{d \beta} \langle Q_L^2 \rangle &=& \frac{d}{d \beta}
\left( Z^2(\beta) \chi a^4(\beta) V + M(\beta) V\right) = 
 \chi V \frac{d}{d \beta} ( Z^2(\beta) a^4(\beta) ) + 
V \frac{d M(\beta)}{d \beta} =  \nonumber \\
&=& \langle Q^2 \rangle a^{-4}(\beta) \frac{d}{d \beta} 
( Z^2(\beta) a^4(\beta) ) + V \frac{d M(\beta)}{d \beta} \; ,
\eeq
which is again consistent with Eq. (\ref{ren-rule}).

This does not complete the discussion, as operators
of higher dimension in the OPE of $q_L(x_3) q_L(x_4)$
could bring corrections in which cannot
be expressed in the same simple form as in Eq. (\ref{ren-rule}).
However we will assume those corrections to be negligible,
and this assumption will be well supported by the numerical
data shown in Section 3.

In order to extract the value of $\langle Q^4 \rangle$ from 
Eq.~(\ref{ren-rule}), we need to 
measure the lattice expectation value $\langle Q_L^4 \rangle$, to 
already know $\langle Q^2 \rangle$, 
and to determine the new renormalization constants
$M_{4,2}(\beta)$ and $M_{4,0}(\beta)$.
The computation of the renormalization constants can be performed using
a simple extension of the heating method. Indeed, measuring the expectation
value $\langle Q_L^4 \rangle$ on the ensemble 
thermalized around a semi-classical configuration of charge $Q$, one obtains
\beq
\langle Q_L^4 \rangle = Z^4(\beta) \; Q^4  + 
M_{4,2}(\beta) \; Q^2 + M_{4,0}(\beta) \; . 
\label{heat-system}
\eeq
It is clear that if we repeat the measurement in 2  sectors with
different values of $Q$ (for instance $Q = 0,1$), 
we obtain 2 different constraints involving
the renormalization constants which, assuming that
$Z(\beta)$ is already known, allow the determination of 
$M_{4,2}(\beta)$ and $M_{4,0}(\beta)$.
If we perform the measurement in more than 2 sectors we 
have more constraints than constants to be determined, and
this offers the possibility to perform a non-trivial test of 
Eq.~(\ref{ren-rule}) and of the heating method.

We close this Section by considering the general case of the 
$n$-th order correlation function. A natural extension of 
Eq.~(\ref{ren-rule}) is the following:
\beq
\langle Q_L^n \rangle = Z(\beta)^n \langle Q^n \rangle + 
\sum_{h=1}^{n/2} 
M_{n,n-2h}(\beta) \langle Q^{n-2h} \rangle \; ,
\label{ren-rule2}
\eeq
and its validity can be discussed along the same lines as for $n = 4$.
In this case, the measurement of $\langle Q_L^n \rangle$ 
on the ensemble thermalized around a semi-classical 
configuration of charge $Q$ gives
\beq
\langle Q_L^n \rangle = Z^n Q^n  + 
M_{n,n-2} Q^{n-2} + \dots + M_{n,0} \; , 
\label{heat-system2}
\eeq
so that it is necessary to measure $\langle Q_L^n \rangle$ in at least
$n/2$ different topological sectors to determine the $n/2$ renormalization
constants $M_{n,n-2}(\beta), M_{n,n-4}(\beta), \; \dots \; , M_{n,0}(\beta)$.

\newsection{Determination of $\langle Q^4 \rangle_c$ in $SU(3)$ pure gauge theory}

In this Section we present numerical results obtained
for $SU(3)$ pure gauge theory. We will illustrate a detailed
study of the renormalization constants 
involved in the determination of $\langle Q^2 \rangle$ and 
$\langle Q^4 \rangle$, using the heating method: 
we will employ samples of configurations
thermalized in three different topological sectors, $Q = 0,1,2$,
and this will enable us to perform a non-trivial consistency test of 
Eq.~(\ref{ren-rule}) and of the heating method itself.
We will then combine the values of the renormalization constants with the 
results obtained at equilibrium to compute $\langle Q^4 \rangle_c$
and thus obtain information about the quartic term in the expansion
of $F(\theta)$. In particular we will compute the quantity
$b_2 = - \frac{\langle Q^4 \rangle_c}{12 \langle Q^2 \rangle}$, 
which measures the relative weight of quartic to quadratic terms and has 
been determined in Ref.~\cite{ehl-02} via the cooling method\footnote{
We use the same notation used in Ref.~\cite{ehl-02}, in order to 
make the comparison easier.
}.

All the results reported in this Section refer to simulations
performed on a lattice of size $16^4$, with $\beta = 6.1$ and 
the standard Wilson action. Two different discretized 
topological charge density operators have been used, corresponding
to the 1-smeared and 2-smeared operators defined in Section 2.

\newsubsection{Determination of the renormalization constants}

We will make use of the method described in Section 2
to determine the renormalization constants which enters the computation
of $\langle Q^2 \rangle$ and $\langle Q^4 \rangle$. 

We have collected five different samples of configurations,
one thermalized in the $Q = 0$ sector (around the zero field
configuration), two in the $Q = 1$ sector (thermalized around two different
semiclassical configurations of topological charge one) and two
in the $Q = 2$ sector (thermalized around two different
semiclassical configurations of topological charge two).
The semiclassical configurations have been obtained by extracting thermalized
configurations with non-trivial topology 
from the equilibrium ensemble at $\beta = 6.1$ and then minimizing
their action by a usual cooling technique.
All the five samples have been obtained by performing about 3000
heating trajectories around the semiclassical configurations, each
trajectory consisting of 90 heating steps; 
6 straight cooling steps have been applied on heated configurations
to check that their background topological content did not change.

We have then measured the expectation values
$\langle Q_L^2 \rangle$, $\langle Q_L^4 \rangle$, 
and also $\langle Q_L \rangle/Q$ where $Q \neq 0$,  over the 
five samples, with the aim of applying Eqs.~(\ref{heat-z}),
(\ref{heat-0}), (\ref{heat-system})
and determine the renormalization constants\footnote{
The initial charge of the semiclassical configuration
used in the heating procedure, $Q$, is never strictly
an integer (apart from the case $Q = 0$). The  
deviation from the corresponding integer value is always around
3\% in our case (e.g., when $Q \sim 1$, we actually have
$Q \simeq 0.97$). When applying Eqs.~(\ref{heat-z}), (\ref{heat-0})
and (\ref{heat-system}), one has to be careful and make use of the real 
value of $Q$ instead of the closest integer. 
For a more accurate discussion
on this point see for instance Ref.~\cite{abf-96}. 
}.
We have reported the results in Table I for the 1-smeared operator 
and in Table II for the 2-smeared operator: 
expectation values obtained on samples with the same $Q$ turned
out to be equal within errors, as they should, and we have reported
in the tables only their weighted averages. It is still possible
to appreciate in Table I and II the agreement for the values
of $Z$ determined in the $Q = 1$ and $Q = 2$ sectors.
We have also reported results for $\langle Q_L^6 \rangle$, which
will be used in Section 4.

Let us rewrite Eq.~(\ref{renchi}) in the form $\langle Q_L^2 \rangle = 
Z(\beta)^2 \langle Q^2 \rangle + M_{2,0}(\beta)$, with 
$M_{2,0}(\beta) = V M(\beta)$.
The information contained in $\langle Q_L^2 \rangle$ for each
different topological sector $Q$ can be used to determine
$M_{2,0}$, using Eq.~(\ref{heat-0}). We have
2 constants to be determined, $Z$ and $M_{2,0}$, and
3 equations ($Q = 0,1,2$), plus two direct determinations
of $Z$ from $\langle Q_L \rangle/Q$, so that there are 5
constraints to be satisfied and only 2 variables to be determined.
The fact that this can be done consistently is a non-trivial
test of the method, and more precisely of the fact that the UV fluctuations
which are responsible for the renormalization constants are decoupled 
from the background topological content, an assumption that is at 
the very basis of the heating method, as it has been explained in Section 2.
In practice we have used the 5 values measured for
$\langle Q_L^2 \rangle$ and  $\langle Q_L \rangle/Q$
to perform a best fit to Eqs.~(\ref{heat-z}) and (\ref{heat-0}),
obtaining best fit values which are reported in Table III
and good $\chi^2/{\rm d.o.f.}$ values
($\sim 0.1$ for the 1-smeared operator and $\sim 0.2$ 
for the 2-smeared operator\footnote{ 
The low values obtained for $\chi^2/{\rm d.o.f.}$ 
can be related  to the fact that the measurements of $\langle Q_L^2 \rangle$ 
and $\langle Q_L \rangle/Q$ at corresponding values of $Q$ are partially 
correlated, since they are measured on the same sample of configurations.
}). The values obtained for $Z$ and $M_{2,0}$ are compatible with those
reported in Ref.~\cite{add-97}.

The same procedure has been repeated, using the $\langle Q_L^4 \rangle$ 
measurements and Eq.~(\ref{heat-system}),
to obtain the best fit values for $M_{4,2}$ and $M_{4,0}$ reported
in Table III. Also in this case we have obtained
good values for $\chi^2/{\rm d.o.f.}$ ($\sim 0.1$ for both 
the 1-smeared and the 2-smeared operator). This represents a strong
numerical support to the validity of Eq.~(\ref{ren-rule}).

\newsubsection{Determination of $\langle Q^2 \rangle$, $\langle Q^4 \rangle$,
and $\langle Q^4 \rangle_c$}

Now that we have determined the renormalization constants we can proceed
to determine $\langle Q^2 \rangle$ and $\langle Q^4 \rangle$.
The equilibrium values for $\langle Q_L^2 \rangle$ and 
$\langle Q_L^4 \rangle$, which are reported in Table IV, have been measured
on a sample of 300K configurations separated by five updating cycles,
each composed of a mixture of 4 over-relaxation + 1 heat-bath updating sweeps;
the reported errors have been estimated by a standard blocking technique.

The value of $\langle Q^2 \rangle$ can be computed  as
\beq
\langle Q^2 \rangle = \frac{\langle Q_L^2 \rangle - M_{2,0}}{Z^2} \; .
\eeq
The expression for $\langle Q^4 \rangle$ follows from Eq.~(\ref{ren-rule}):
\beq
\langle Q^4 \rangle = \frac{\langle Q_L^4 \rangle - 
M_{4,2} \langle Q^2 \rangle - M_{4,0}}{Z^4} \; .
\eeq
The results for  $\langle Q^2 \rangle$ and  $\langle Q^4 \rangle$
are reported in Table IV, the errors have been computed by 
standard error propagation. It is interesting to notice that the
values obtained for the 1-smeared operator and for the 2-smeared operator
are in good agreement, as they should, again confirming the 
robustness of the method.
We can finally determine $\langle Q^4 \rangle_c = 
\langle Q^4 \rangle - 3 \langle Q^2 \rangle^2$, obtaining
$\langle Q^4 \rangle_c = 0.32 \pm 1.80$ for the 1-smeared and
$\langle Q^4 \rangle_c = 0.66 \; \pm \; 0.90$ for the 2-smeared operator,
leading to $b_2 = -0.012(62)$ and $b_2 = -0.024(32)$ for the 
1-smeared and 2-smeared operator respectively, 
in agreement with the determination reported in Ref.~\cite{ehl-02}.

\newsection{A closer look into the renormalization effects}

The renormalization constants $Z$, $M_{n,m}$ ($m < n$) which, for a 
given lattice discretization $Q_L$, appear in Eq.~(\ref{ren-rule2}),
are in principle independent of each other, or at least no simple
relation exists among them, unless some further hypothesis 
can be done about the nature of the UV fluctuations which are responsible for
the renormalizations. In this Section we will propose and test
an ansatz which will greatly simplify the structure of the renormalization
constants and will lead to a renormalization formula which directly 
involves the connected  correlation functions, thus allowing a more 
precise determination of $b_2$.

An hypothesis about the nature of the UV fluctuations has been done in 
Refs.~\cite{teper-89,dv-92}, where it was assumed that the 
discretized topological charge density can be expressed as
\beq
q_L(x) \simeq [Z + \zeta (x)] q(x) + \eta (x) \; ,
\label{hypo1}
\eeq
where $q(x)$ is a background topological charge density
which is determined by physical fluctuations on the scale
of the correlation length $\xi$, whereas $\zeta(x)$ and $\eta(x)$ 
are random variables with zero averages which are determined
by the short range UV fluctuations and, at least in the continuum
limit, are expected to be decoupled from $q(x)$, i.e.
$\langle \zeta(x) q(x) \rangle = \langle \eta(x) q(x) \rangle = 0$.
Summing Eq.~(\ref{hypo1}) over all lattice points, the following
relation follows for the lattice topological charge $Q_L$:
\beq
Q_L = Z \; Q + \sum_x \zeta(x) q(x) + \eta \; ,
\label{hypo2}
\eeq
where $\eta =  \sum_x \eta(x)$. We now make the further assumption that
the term $\sum_x \zeta(x) q(x)$ in Eq.~(\ref{hypo2}) can be neglected, 
configuration by configuration. This is not  
unreasonable, in view of the fact that $q(x)$ and $\zeta (x)$ are
decoupled from each other.
We will thus assume that 
\beq
Q_L =  Z \; Q + \eta \; , 
\label{ansatz}
\eeq
where $\eta$ is a random noise with zero average which is 
stochastically independent of $Q$. 

This assumption has relevant consequences for the structure of the 
renormalization constants. Indeed, using the hypothesis that 
$Q$ and $\eta$ are stochastically independent variables and that they are
both evenly distributed around zero, it is easy to verify that the 
general renormalization formula in Eq.~(\ref{ren-rule2}) becomes
\beq
\langle Q_L^n \rangle = \sum_{h = 0}^{n/2} {n \choose 2h} Z^{n - 2h}  
\langle Q^{n - 2h} \rangle \langle \eta^{2h} \rangle \; ,
\label{ren-rule-2}
\eeq
so that the renormalization relation for $\langle Q_L^n \rangle$
is described only in terms of $Z$ and of the correlation
functions of the noise $\eta$. In particular we have
$M_{n,m} = {n \choose m} Z^m \langle \eta^{n-m} \rangle$,
a relation that should be verified on numerical data if our ansatz in 
Eq.~(\ref{ansatz}) is correct. From the data in Table III it can 
be checked that indeed $M_{4,2} = 6 Z^2 \langle \eta^2 \rangle = 
6 Z^2 M_{2,0}$, but we will now proceed further and check
the validity of Eq.~(\ref{ren-rule-2}) up to $n = 6$.
The correlation functions of $\eta$ can be determined by the heating
method using the analogous of Eq.~(\ref{heat-system2}), which
up to $n = 6$ reads:
\beq
\langle Q_L^2 \rangle &=& Z^2 Q^2 + \langle \eta^2 \rangle
\nonumber \\
\langle Q_L^4 \rangle &=& Z^4 Q^4  + 
6 Z^2 Q^2  \langle \eta^2 \rangle + \langle \eta^4 \rangle
\nonumber \\
\langle Q_L^6 \rangle &=& Z^6 Q^6 + 
15 Z^4 Q^4  \langle \eta^2 \rangle + 
15 Z^2 Q^2  \langle \eta^4 \rangle + \langle \eta^6 \rangle \; .
\label{heat-system-2}
\eeq
Using the values for $\langle Q_L^2 \rangle$, $\langle Q_L^4 \rangle$ and
$\langle Q_L^6 \rangle$ obtained in the sectors with $Q = 0,1,2$ and
reported in Tables I and II, we have performed a best fit to 
Eqs.~(\ref{heat-system-2}), obtaining the best fit values reported in Table V 
with  $\chi^2/{\rm d.o.f.} \simeq 0.34$ for the 1-smeared operator and 
$\chi^2/{\rm d.o.f.} \simeq 0.23$ for 2-smeared operator.
The fact that the values for 
$\langle Q_L^2 \rangle$, $\langle Q_L^4 \rangle$ and
$\langle Q_L^6 \rangle$ obtained in the various 
sectors can be fitted by the simple relations in Eq.~(\ref{heat-system-2})
is a confirmation of the validity of the ansatz 
in Eq.~(\ref{ansatz}).

Assuming that Eq.~(\ref{ansatz}) is valid, it is 
possible to write a renormalization relation which involves directly
the connected correlation functions. Indeed, it is a general rule
that the connected correlation functions of a stochastic variable
($Q_L$ in our case), which is the sum of two variables which are 
stochastically independent of each other ($Z Q$ and $\eta$ in our case),
are the sum of the corresponding connected correlation functions, i.e.
\beq
\langle Q_L^n \rangle_c = Z^n \langle Q^n \rangle_c + 
\langle \eta^n \rangle_c \; .
\label{simple-ren}
\eeq

Therefore in order to compute $\langle Q^n \rangle_c$
we need to know, apart from $Z$, only one renormalization constant,
$\langle \eta^n \rangle_c$, which can be easily measured 
by computing $\langle Q_L^n \rangle_c$ on the sample of configurations
in the $Q = 0$ sector. 
$\langle Q^n \rangle_c$ is then given by
\beq
\langle Q^n \rangle_c = 
\frac{\langle Q_L^n \rangle_c - \langle \eta^n \rangle_c}{Z^n} \; ,
\label{simple}
\eeq
where $\langle Q_L^n \rangle_c$ is measured on the ensemble of 
configurations at equilibrium.

We have computed 
$\langle Q_L^4 \rangle_c = \langle Q_L^4 \rangle - 3 \langle Q_L^2 \rangle^2 $ 
on our equilibrium configurations at $\beta = 6.1$, obtaining  
$\langle Q_L^4 \rangle_c = 0.026(7)$ for the 1-smeared operator
and $\langle Q_L^4 \rangle_c = 0.057(13)$ for the 2-smeared operator.
We have then computed $\langle Q_L^4 \rangle_c$ on our sample of configurations
thermalized in the $Q = 0$ sector at $\beta = 6.1$ , obtaining
$\langle \eta^4 \rangle_c = \langle \eta^4 \rangle - 3 \langle \eta^2 \rangle^2 = 0.006(4)$ for the 1-smeared operator
and $\langle \eta^4 \rangle_c = 0.001(2)$ for the 2-smeared operator.
In both cases (equilibrium and $Q = 0$) errors have been estimated by standard
 jackknife techniques.

By using Eq.~(\ref{simple}) and the values for $Z$ and $\langle Q^2 \rangle$
previously obtained, we have obtained $\langle Q^4 \rangle_c = 0.68(24)$,
$b_2 = -0.024(9)$ for the 1-smeared operator and 
$\langle Q^4 \rangle_c = 0.66(15)$, $b_2 = -0.024(6)$ 
for the 2-smeared operator.

By making use of the ansatz in Eq.~(\ref{ansatz}) we have thus
made determinations which are much more precise than those obtained
in Section 3. The reason is that Eq.~(\ref{simple-ren}) allows
to relate $\langle Q^n \rangle_c$ directly to the connected correlation
functions of the discretized lattice topological charge, with only two
renormalization constants involved: this greatly simplifies
computations and error propagation, thus leading to improved
estimates. 
We notice that most of the error in the final
determination of $\langle Q^4 \rangle_c$ and  $b_2$ comes
from the determination of $\langle Q_L^4 \rangle_c$ at equilibrium,
which is also the most expensive part of the computation in terms
of CPU time. The renormalization procedure is thus completely under
control and numerically non expensive.

We have also made a determination of $b_2$ at $\beta = 6.0$,
again on a $16^4$ lattice.
On a sample of about 300K configurations obtained at equilibrium and
using the same algorithm as for $\beta = 6.1$ we have obtained,
for the 2-smeared operator, $\langle Q_L^2 \rangle = 1.377(7)$,  
$\langle Q_L^4 \rangle_c = 0.052(23)$. On a sample of configurations
thermalized in the $Q = 0$ topological sector by performing about
3000 heating trajectories, each composed of 90 heating steps, we
have obtained, for the 2-smeared operator,  
$\langle \eta^2 \rangle = 0.308(10)$ and 
$\langle \eta^4 \rangle_c = 0.002(3)$. From these data, using 
the value $Z(\beta = 6.0) = 0.51(2)$ reported for the 2-smeared operator
in Ref.~\cite{add-97}, we obtain $b_2 = -0.015(8)$, which is consistent
with the value obtained at $\beta = 6.1$.

Let us close this Section with some further speculations about the nature
of the UV fluctuations. The value obtained for $\langle \eta^4 \rangle_c$
is very small and compatible with zero for both the 1-smeared and 
the 2-smeared operator. We have also measured $\langle \eta^6 \rangle_c$
 on the sample of configurations at $Q = 0$ obtaining
$\langle \eta^6 \rangle_c = 0.001(8)$ for the 1-smeared and
$\langle \eta^6 \rangle_c = 0.0005(14)$ for 2-smeared operator ($\beta = 6.1$).
This seems to indicate that $\eta$ behaves as a Gaussian noise:
it can also be verified on the values reported in Table V
that $\langle \eta^4 \rangle$ is always compatible with
$3 \langle \eta^2 \rangle^2$ and that $\langle \eta^6 \rangle$ is 
always compatible with $15 \langle \eta^2 \rangle^3$, as it should
be for a Gaussian variable. We have also directly verified on 
the histograms of the distribution of $\eta$ in the $Q = 0$ sector
that it does not present any significant deviation from 
a Gaussian distribution. If the hypothesis of Gaussian distribution
for $\eta$ were true, the renormalization relation in Eq.~(\ref{simple-ren})
would become, for  $n > 2$,
$\langle Q_L^n \rangle_c = Z^n \langle Q^n \rangle_c$, 
i.e. the connected correlation functions of $Q_L$ would renormalize
only multiplicatively for $n > 2$. However we will not consider 
further analysis of this hypothesis in the present paper.

\newsection{Conclusions}

In this paper we have discussed how to extend
the field theoretical method, already used for
the lattice determination of the topological 
susceptibility, in order to compute
further terms in the expansion of the 
ground state energy $F(\theta)$ around $\theta = 0$.

After a review of the method as used for the 
determination of the topological susceptibility,
we have discussed the structure of 
the renormalizations involved in the general case
and how they can be computed using the heating method.
We have presented numerical results regarding $SU(3)$
pure gauge theory, providing both support to
the correctness of the method and a determination
of the fourth order term in the expansion of 
$F(\theta)$ around $\theta = 0$.

In the last part of the paper we have made some speculations
about the nature of the lattice UV fluctuations which are responsible for
the renormalizations and proposed an ansatz, which we have verified
against the numerical data, that leads to a simpler structure
for the renormalizations and to an improved estimate of 
the fourth order term.

The determinations obtained are in agreement with those
obtained in Ref.~\cite{ehl-02} via the cooling method,
and  confirm that fourth order corrections to  
the simple $\theta^2$ behaviour of $F(\theta)$ around $\theta = 0$
are small already for $N = 3$.

\bigskip
\noindent {\bf Acknowledgements}
\smallskip

We thank B.~All\'es, L.~Del~Debbio, A.~Di~Giacomo and E.~Vicari for 
useful comments and discussions.
This work has been partially supported by MIUR.  
We thank the computer center of ENEA for providing us with time on
their QUADRICS machines.

\vfill\eject

{\renewcommand{\Large}{\normalsize}
}

\vfill\eject

\noindent
\begin{center}
{\bf TABLE CAPTIONS}
\end{center}
\vskip 0.5 cm
\begin{itemize}
\item [\bf Tab.~I.] Expectation values measured in different 
topological sectors for the 1-smeared operator.
\bigskip
\item [\bf Tab.~II.] Expectation values measured in different 
topological sectors for the 2-smeared operator.
\bigskip
\item [\bf Tab.~III.] Values of the renormalization constants
obtained by using the results reported in Tables I and II and
performing a best fit to Eqs.~(\ref{heat-z}), (\ref{heat-0})
and (\ref{heat-system}).
\bigskip
\item [\bf Tab.~IV.] Expectation values measured at eequilibrium
and results obtained for the renormalized quantities.
\bigskip
\item [\bf Tab.~V.] 
Values of the renormalization constants
obtained by using the results reported in Tables I and II and
performing a best fit to Eqs.~(\ref{heat-z}) and
(\ref{heat-system-2}).
\bigskip
\end{itemize}

\vfill\eject

\vskip 9mm

\centerline{\bf Table I}

\vskip 4mm

\moveright 0.70 in
\vbox{\offinterlineskip
\halign{\strut
\vrule \hfil\quad $#$ \hfil \quad & 
\vrule \hfil\quad $#$ \hfil \quad & 
\vrule \hfil\quad $#$ \hfil \quad & 
\vrule \hfil\quad $#$ \hfil \quad & 
\vrule \hfil\quad $#$ \hfil \quad \vrule \cr
\noalign{\hrule}
Q & Z = \langle Q_L \rangle/Q & \langle Q_L^2 \rangle & 
\langle Q_L^4 \rangle & \langle Q_L^6 \rangle \cr
 & & & & \cr
\noalign{\hrule}
\noalign{\hrule}
0 & -   & 0.311(12)  & 0.290(20) & 0.48(5) \cr 
\noalign{\hrule}
1 & 0.416(6) & 0.4785(60) & 0.630(15) & 1.295(45) \cr 
\noalign{\hrule}
2 & 0.413(5) & 0.9626(80) & 1.973(50) & 5.82(18) \cr
\noalign{\hrule}
}}

\vskip 9mm

\centerline{\bf Table II}

\vskip 4mm

\moveright 0.72 in
\vbox{\offinterlineskip
\halign{\strut
\vrule \hfil\quad $#$ \hfil \quad & 
\vrule \hfil\quad $#$ \hfil \quad & 
\vrule \hfil\quad $#$ \hfil \quad & 
\vrule \hfil\quad $#$ \hfil \quad & 
\vrule \hfil\quad $#$ \hfil \quad \vrule \cr
\noalign{\hrule}
Q & Z = \langle Q_L \rangle/Q & \langle Q_L^2 \rangle & 
\langle Q_L^4 \rangle & \langle Q_L^6 \rangle \cr
 & & & & \cr
\noalign{\hrule}
\noalign{\hrule}
0 & - & 0.208(10)& 0.124(10) & 0.131(18) \cr 
\noalign{\hrule}
1 & 0.544(5) & 0.489(5) & 0.556(12) & 0.93(3) \cr 
\noalign{\hrule}
2 & 0.542(4) & 1.314(8) & 2.77(6) & 7.65(17) \cr
\noalign{\hrule}
}}

\vskip 9mm

\centerline{\bf Table III}

\vskip 4mm

\moveright 0.50 in
\vbox{\offinterlineskip
\halign{\strut
\vrule \hfil\quad $#$ \hfil \quad & 
\vrule \hfil\quad $#$ \hfil \quad & 
\vrule \hfil\quad $#$ \hfil \quad & 
\vrule \hfil\quad $#$ \hfil \quad & 
\vrule \hfil\quad $#$ \hfil \quad \vrule \cr
\noalign{\hrule}
{\rm operator} & Z & M_{2,0} & M_{4,2} & M_{4,0} \cr
 & & & & \cr
\noalign{\hrule}
\noalign{\hrule}
{\rm 1-smeared} &  0.414(4) & 0.315(6) & 0.336(16) & 0.289(16) \cr 
\noalign{\hrule}
{\rm 2-smeared} &  0.543(5) & 0.211(5) & 0.377(15) & 0.124(9) \cr
\noalign{\hrule}
}}

\vskip 9mm

\centerline{\bf Table IV}

\vskip 4mm

\moveright 0.36 in
\vbox{\offinterlineskip
\halign{\strut
\vrule \hfil\quad $#$ \hfil \quad & 
\vrule \hfil\quad $#$ \hfil \quad & 
\vrule \hfil\quad $#$ \hfil \quad & 
\vrule \hfil\quad $#$ \hfil \quad & 
\vrule \hfil\quad $#$ \hfil \quad \vrule \cr
\noalign{\hrule}
{\rm operator} & \langle Q_L^2 \rangle & \langle Q_L^4 \rangle  & \langle Q^2 \rangle & \langle Q^4 \rangle  \cr
 & & & & \cr
\noalign{\hrule}
\noalign{\hrule}
{\rm 1-smeared} & 0.7121(38)  & 1.548(18) & 2.312(72) & 16.4 \pm 1.8 \cr 
\noalign{\hrule}
{\rm 2-smeared} &  0.8776(60) & 2.368(36) & 2.262(41) & 16.02(72) \cr
\noalign{\hrule}
}}

\vskip 9mm

\centerline{\bf Table V}

\vskip 4mm

\moveright 0.50 in
\vbox{\offinterlineskip
\halign{\strut
\vrule \hfil\quad $#$ \hfil \quad & 
\vrule \hfil\quad $#$ \hfil \quad & 
\vrule \hfil\quad $#$ \hfil \quad & 
\vrule \hfil\quad $#$ \hfil \quad & 
\vrule \hfil\quad $#$ \hfil \quad \vrule \cr
\noalign{\hrule}
{\rm operator} & Z & \langle \eta^2 \rangle & \langle \eta^4 \rangle & \langle \eta^6 \rangle \cr
 & & & & \cr
\noalign{\hrule}
\noalign{\hrule}
{\rm 1-smeared} &  0.415(4) & 0.315(6) & 0.298(11) & 0.462(37) \cr 
\noalign{\hrule}
{\rm 2-smeared} & 0.542(4) & 0.211(5) & 0.129(6) & 0.131(17)  \cr
\noalign{\hrule}
}}

\end{document}